\documentclass[%
 aip,
 amsmath,amssymb,
 reprint,%
]{revtex4-1}

\usepackage{graphicx}% Include figure files
\usepackage{dcolumn}% Align table columns on decimal point
\usepackage{bm}% bold math
\usepackage[utf8]{inputenc}
\usepackage[T1]{fontenc}
\usepackage{mathptmx}
\usepackage{etoolbox}
\usepackage{empheq}
\usepackage{amsfonts}
\usepackage{amssymb}
\usepackage{mathtools}
\usepackage{tensor}
\usepackage{numprint}
\usepackage{siunitx}
\usepackage{subfigure}
\usepackage{enumitem}
\usepackage{stackengine}
\usepackage{natbib}
\usepackage[version=4]{mhchem} % Formula subscripts using \ce{}
\usepackage[dvipsnames]{xcolor}
\usepackage{amsmath}
\newcommand{\bss}{\boldsymbol{s}}
\newcommand{\bR}{\boldsymbol{R}}

\newcommand{\oxchem}{Physical and Theoretical Chemistry Laboratory, Department of Chemistry, University of Oxford, Oxford OX1 3QZ, United Kingdom.}
\newcommand{\hwu}{Institute of Chemical Sciences, School of Engineering and Physical Sciences, Heriot-Watt University, Edinburgh EH14 4AS, United Kingdom.}

\renewcommand{\Re}{\operatorname{Re}}

\makeatletter
\def\@email#1#2{%
 \endgroup
 \patchcmd{\titleblock@produce}
  {\frontmatter@RRAPformat}
  {\frontmatter@RRAPformat{\produce@RRAP{*#1\href{mailto:#2}{#2}}}\frontmatter@RRAPformat}
  {}{}
}%
\makeatother

\begin{document}

\preprint{AIP/123-QED}

%\title[Multistate MASH]{Predicting the Ultrafast electron diffraction of gas-phase cyclobutanone using multi-state MASH}
\title[Multistate MASH]{Using a multistate Mapping Approach to Surface Hopping to predict the Ultrafast Electron Diffraction signal of gas-phase cyclobutanone}
% Force line breaks with \\

% I just put names in as placeholders - discuss final list and order with Lewis and Andres before sending out
\author{Lewis Hutton}\email{lewis.hutton@chem.ox.ac.uk}
\affiliation{\oxchem}
\author{Andr\'{e}s Moreno Carrascosa}
\affiliation{\oxchem}
\author{Andrew W. Prentice}
\affiliation{\hwu}
\author{Mats Simmermacher}
\affiliation{\oxchem}
\author{Johan E.\ Runeson}
\affiliation{\oxchem}
\author{Martin J.\ Paterson}
\affiliation{\hwu}
\author{Adam Kirrander}\email{adam.kirrander@chem.ox.ac.uk}
\affiliation{\oxchem}

\date{\today}% It is always \today, today,

\begin{abstract}
%Using the newly developed multistate Mapping Approach to Surface Hopping (mulitstate MASH) method coupled with SA(3)-CASSCF(12,12), the gas-phase Ultrafast Electron Diffraction (UED) of cyclobutanone has been predicted and analyzed as part of the cyclobutanone prediction challenge. After excitation into the n-3s Rydberg state (S$_2$), the cyclobutanone molecule can deactivate through several pathways, including $\mathrm{\alpha}$ ring-opening, ethene/ketene production, and \ce{CO} liberation. With the aid of time-resolved UED signals, the topology of these pathways and the molecular geometries involved have been characterized, giving rise to four possible outcomes and agreeing with previous findings.

Using the recently developed multistate mapping approach to surface hopping (multistate MASH) method combined with SA(3)-CASSCF(12,12)/aug-cc-pVDZ electronic structure calculations, the gas-phase isotropic ultrafast electron diffraction (UED) of cyclobutanone is predicted and analyzed. After excitation into the n-3s Rydberg state (S$_2$), cyclobutanone can relax through two S$_2$/S$_1$ conical intersections, one characterized by compression of the \ce{CO} bond, the other by dissociation of the $\mathrm{\alpha}$-CC bond. Subsequent transfer into the ground state (S$_0$) is then achieved via two additional S$_1$/S$_0$ conical intersections that lead to three reaction pathways: $\mathrm{\alpha}$ ring-opening, ethene/ketene production, and \ce{CO} liberation. The isotropic gas-phase UED signal is predicted from the multistate MASH simulations, allowing for a direct comparison to experimental data. This work, which is a contribution to the cyclobutanone prediction challenge, facilitates the identification of the main photoproducts in the UED signal and thereby emphasizes the importance of dynamics simulations for the interpretation of ultrafast experiments.

\end{abstract}

\maketitle

\section{\label{sec:introduction} Introduction}

%\subsection{\label{Cyclobutanone} Cyclobutanone}
%%%%%%%%%%%%%
\par The photochemistry of cyclic ketones has been studied extensively.\cite{Benson1942,  Lee1971, Baba1984, Tang1976, Harrison1980, Causley1980} Despite their small size, they have a rich photochemistry with multiple competing pathways that include dissociation, fluorescence, and intersystem crossing.\cite{Kao2020, Hemminger1972} The relative importance of the different pathways is closely linked to the size of the organic ring and the associated ring strain.\cite{Kao2020} The photodissociation of cyclic ketones has been of particular interest ever since the pioneering work of Norrish.\cite{Saltmarsh1935} 

%Further pathways producing \ce{CH_2CH_2} + \ce{CH_2=CO} and \ce{CH_2CH_2CH_2} + \ce{CO} have also been observed.\cite{Benson1942, Diau2001}
\par A target of recurring interest in this class of molecules is cyclobutanone, \ce{(CH2)3CO}.\cite{Benson1942, Denschlag1968, Lee1969, Lee1971, Diau2001, Kao2020} Early experimental studies identified the photoproducts of cyclobutanone as propylene (cyclopropane) and carbon monoxide (\ce{C_3H_6} + \ce{CO}) or ethylene and ketene (\ce{CH2CH2} + \ce{CH2CO}). The corresponding quantum yields were found to be 40\% and 60\%,\cite{Benson1942} although these quantum yields were later shown to have a strong dependence on the excitation wavelength.\cite{Hemminger1972} Recent experimental work includes ultrafast transient absorption spectroscopy on cyclobutanone in solution.\cite{Kao2020} Upon excitation to the S$_1$ (n$\pi^*$) state using UV pulses in the range 255–312 nm UV pulses, Kao \emph{et al.} found that singlet dissociation pathways are dominant at shorter wavelengths, predominantly via Norrish Type-1 $\alpha$ cleavage, with minor contributions from ketene formation. At higher excitation energies, such as 200 nm, the n-3s Rydberg state comes into play.\cite{Drury-Lessard1978, OToole1991, Whitlock1971} In this regime, using time-resolved mass spectrometry and photoelectron spectroscopy, Kuhlman \emph{et al.} found that a ring puckering mode effectively couples the S$_2$ and S$_1$ states, allowing for rapid internal conversion.\cite{Kuhlman2012}

\par The experimental studies have been accompanied by theoretical work. Electronic structure calculations using complete active space self-consistent field, CASSCF(10,8), and multistate complete active space second-order perturbation theory, MS-CASPT2(10,8), predicted a small barrier for ring-opening on the S$_1$ state, with the possibility of minor contributions from barrierless dissociation in the T$_1$ state.\cite{Xia2015a} Building on this, simulations of cyclobutanone by \emph{ab initio} multiple spawning (AIMS) found a ring-opening mechanism to be dominant with a small portion of trajectories producing \ce{CH_2=CH_2} + \ce{CH_2CO} and \ce{CH_2CH_2} + \ce{CH2=CO} fragmentation.\cite{Liu2016}  We also note that Kuhlman \emph{et al.} constructed a five-dimensional linear-vibronic Hamiltonian (LVH) model to simulate the S$_2$/S$_1$ decay in cyclobutanone using the multiconfigurational time-dependent Hartree (MCTDH) method, which succeeded in replicating the short time constants of 0.95 ps. \cite{Kuhlman2012a}
%\YYY{[Lewis, as far as I can tell  you have written the same set of photoproducts twice!]}

\par The current paper is motivated the prediction challenge based on an UED experiment carried out at the SLAC MeV-UED facility.\cite{prediction-challenge2024} Ultrafast electron diffraction has emerged as a powerful experimental technique for observing molecular dynamics in the recent decade,\cite{Miller2015,Centurion2015,Yang2018,Wolf2019,RazmusPCCP2022} alongside ultrafast x-ray scattering. \cite{Minitti2015,Ruddock2019,Ruddock2019SciAdv,GabalskiJCP2022} Notably, both techniques have made advances in extracting information that extends beyond structural dynamics.\cite{Yang2020,Yong2020} We use this challenge as an opportunity to benchmark the electronic structure calculations for cyclobutanone. This comparison includes non-standard selected-CI electronic structure methods that overcome the issue of large active space selection, constitute a black-box alternative to complete active space (CAS) methods and provide complementary benchmarks.\cite{Paterson2023, Jacquemin2023} The nonadiabatic simulations are carried out using the newly developed multistate MASH method.\cite{Runeson2023a} Furthermore, we use the simulations to predict the UED signals and analyze these to determine the main contributions to the scattering.

\section{\label{sec:theory}Theoretical methods}
\subsection{\label{Molecular_dynamics} Nonadiabatic Molecular dynamics}

\par Many computational methods have been developed to simulate nonadiabatic dynamics over the past decades. Some, such as MCTDH \cite{Meyer1990, Beck2000c} and Multi-Layer MCTDH,\cite{Wang2005} can provide numerically exact results. However, these methods require high-accuracy precomputed potential energy surfaces (PESs), which, combined with the exponential scaling of quantum mechanics, imposes severe limitations on the number of degrees of freedom that can be treated. To circumvent this problem, the nuclear wavefunction can be represented as a linear combination of traveling Gaussian basis functions or classically-guided trajectories, and the electronic structure quantities needed can be calculated {\emph{on-the-fly}}, also known as direct dynamics. Such approaches can significantly reduce the phase space explored and therefore the computational cost per degree of freedom. {\emph{On-the-fly}} methods include examples where the nuclei are propagated using equations of motion derived from the variational principle, notably variational multiconfigurational Gaussian (vMCG),\cite{Worth2004a, Lasorne2006, Richings2015} or semiclassically, which includes methods such as full or {\emph{ab initio}} multiple spawning (FMS/AIMS)\cite{Curchod2018d, Martinez1997, Martinez1996, Martinez1997a} and multiconfigurational Ehrenfest (MCE),\cite{Shalashilin2009,Saita2012a,Shalashilin2017} or classically, such as trajectory surface hopping (TSH).\cite{Tully1990a} In addition to the above methods, mapping methods treat the electronic and nuclear degrees of freedom in an equal manner. Examples include the Meyer-Miller-Stock-Thoss (MMST) mapping Hamiltonian, symmetrical quasi-classical windowing, and generalized spin mapping.\cite{Meyera1979, Cotton2013, Miller2016, Runeson2020} Ideas from this last category informed the development of the surface-hopping variant used in this paper (see Section \ref{sec:MASH} below). For an in-depth overview of different methods for nonadiabatic dynamics, the reader is directed towards the edited volume in reference \citenum{LeticiaRoland}.

%\par A plethora of methods have been developed and used to simulate nonadiabatic dynamics including full multiple spawning (FMS)/ AIMS, \cite{Curchod2018d, Martinez1997, Martinez1996, Martinez1997a} variational multiconfigurational Gaussian (vMCG), \cite{Worth2004a, Lasorne2006, Richings2015} MCTDH, \cite{Meyer1990, Beck2000c}, multiconfigurational Erhenfest dynamics (MCE), \cite{Shalashilin2009, Saita2012a} and trajectory surface hopping (TSH). \cite{Tully1990a} In addition to these methods, there exists mapping methods, where the electronic and nuclear degrees of freedom are treated equally. Examples of this are Meyer-Miller-Stock-Thoss (MMST) and mapping symmetrical quasi-classical windowing. \cite{Meyera1979, Cotton2013, Cotton2015} Each method has its own advantages and disadvantages such as MCTDH yielding highly accurate results but at the cost of requiring expensive precomputed potential energy surfaces (PESs). For a more in-depth view of different methods of non-adiabatic dynamics, the interested reader is directed towards Ref \citenum{Curchod2018e}. 
\subsubsection{\label{Surface_Hopping} Trajectory surface hopping}

%Erhenfest dynamics having deterministic forces allowing 'better' long-time limit dynamics, however, Erhenfest dynamics can also propagate on inverted potentials with negative populations.  

% However, there also exists methods in the so-called mapping methods such as Ehrenfest dynamics or Meyer-Miller-Stock-Thoss (MMST) mapping. In mapping methods both the nuclear and electronic degrees of freedom are treated on the same footing. Each method has it's own advantages and disadvantages. With Ehrenfest dynamics having been shown to yield better results for system-bath models, however, Erhenfest dynamics can suffer from evolution on inverted potentials when negative populations are observed. 

\par For nonadiabatic molecular dynamics, TSH is perhaps the most commonly employed approach to support the interpretation of experiments. \cite{Ruckenbauer2016,BellshawCPL2017,Squibb2018,Polyak2019a,Spiridoula2020,Merritt2021,Hutton2022a,Rolles2024} In TSH, an ensemble of individual trajectories is used to represent the propagation of the nuclear wavepacket. In each trajectory, the nuclei are treated classically and the electronic amplitudes are used to predict hops to other electronic states. For a full discussion of TSH please see reference \citenum{Barbatti2011}, while we continue by highlighting the key aspects of the fewest switches surface hopping (FSSH) variant of TSH.

\par Each trajectory in FSSH is initialized by projecting a ground-state vibrational distribution into the excited state of interest. We consider pure excitation, \textit{i.e.}\ $c_a = 1$, where $c_a$ is the electronic wavefunction coefficient for the active state $a$.  The nuclear coordinates, $\bar{\mathbf{R}}$, are propagated using classical equations of motion, 
\begin{equation}
    \frac{d^2\mathbf{\bar{R}_{\alpha}}}{dt^{2}} = -\frac{1}{M_{\alpha}} \nabla_{\alpha}E_{a},
\end{equation}
\noindent for each atom $\alpha$, with $t$ the time, $M_{\alpha}$ the mass of atom $\alpha$, and $\nabla_{\alpha}E_{a}$ the gradient (force) on the {\emph{active}} electronic adiabatic state. The integration is typically carried out using a standard velocity-Verlet scheme, though other propagators may be able to achieve better convergence.\cite{BLANES2002} The forces are normally obtained from electronic structure calculations of adiabatic states. In parallel with the nuclear propagation, one must integrate the electronic coefficients, $c=(c_1,\ldots,c_N)$, for the $N$ electronic states considered. In the adiabatic representation, the probability $P^{\mathrm{FSSH}}_{a \rightarrow b}$ of a hop from state $a$ to state $b$, is evaluated according to,
\begin{equation}
\label{eq:SH_hop_prob}
    P^{\mathrm{FSSH}}_{a \rightarrow b} = \mathrm{max}\Bigr[0, \frac{2\Delta t}{|c_a|^2} \Re(\mathbf{d}_{ab} \cdot \mathbf{\bar{v}}  c_b c_a^{*}) \Bigr] \text{,}
\end{equation}
\noindent where $\mathbf{d}_{ab}$ is the nonadiabatic coupling vector between states $a$ and $b$, $\mathbf{\bar{v}}$ is the classical nuclear velocity, $c_a$ and $c_b$ are the electronic coefficients for states $a$ and $b$, and $\Delta t$ is the timestep.\cite{Crespo-Otero2018c} In some cases, an overlap scheme is employed to avoid the expensive calculations of nonadiabatic couplings.\cite{Hammes-Schiffer1994, Plasser2016}  At each time step, $P^{\mathrm{FSSH}}_{a \rightarrow b}$ is evaluated and compared to a random number, resulting in either acceptance or rejection of the hop. If a hop is accepted but the energy in the system is insufficient for it to occur, the hop is rejected and the nuclear velocities are reflected. This scenario is referred to as a frustrated hop. For hops that do take place, the active state is updated ($a\rightarrow b$), the velocities are modified to conserve total energy, and the trajectories continue to be propagated using the forces on the new active state.   

The algorithm presented above is robust, but also causes one of the most well-known shortcomings of TSH, the so-called overcoherence issue. A number of, more or less \textit{ad hoc}, decoherence correction schemes have been developed in response.\cite{Granucci2010,Subotnik2016review} The method presented next aims to alleviate the need for such \emph{ad hoc} corrections.

\subsubsection{\label{sec:MASH} Multistate mapping approach to surface hopping}
\par Recently, a new version of TSH was devised by Mannouch and Richardson, known as "mapping approach to surface hopping" (MASH).\cite{Mannouch2023} MASH combines elements of mapping methods with TSH to attempt to get the advantages of both. Hence, in contrast to the stochastic algorithm in FSSH, MASH uses a deterministic algorithm to evaluate the active state,

\begin{equation} 
P^{\mathrm{MASH}}_{a \rightarrow b} = 
\begin{cases}
    1,& \text{if } |c_{b}|^2 > |c_{a}|^2 \\
    0,& \text{otherwise.} \\
\end{cases}
\end{equation}

\noindent In other words, MASH defines the active state to be the state with the largest value of $|c_i|^2$, alleviating the need for \emph{ad hoc} decoherence corrections and improving accuracy when compared to the original FSSH procedure.\cite{Mannouch2023} Originally, MASH was formulated for two-state systems, but MASH was then generalized to any number of states by Runeson and Manolopoulos, giving rise to what we hereon refer to as multistate MASH.\cite{Runeson2023a, Runeson2024} 

%\YYY{The method replaces the need for ad hoc decoherence corrections by a rigorous quantum-jump procedure, and even without such jumps it has been found to improve accuracy compared to FSSH in a variety of model systems.[Cite 68]}

%MASH naturally takes into account decoherence by never having a mismatch between the active state and the state with the largest coefficient, $\max{\{|c_i|^2\}}$.

\par To start trajectories, multistate MASH determines the nuclear initial conditions in the same manner as TSH, mapping the ground-state wavefunction onto a classical phase space. However, following other mapping approaches, MASH also considers the electronic coefficients $c$ as a phase space variable. Therefore, rather than describing the initial state as pure, such as is done in FSSH, multistate MASH represents the initial state as a distribution over all values of $c=(c_1,\ldots,c_N)$. In practice, this is done by random assignment of the complex-valued coefficients, $c_i=x_i + \mathrm{i} y_i$, for each electronic state from Gaussian distributions with zero mean and standard deviation one for both the real and imaginary components. The resulting set of coefficients is normalized such that $1=\sum_{i=1}^N|c_i|^2$. Coefficients where the initial state of interest does {\emph{not}} have the largest absolute value are resampled until suitable sets are found. It is worth noting that other choices of initial distributions are possible, but have been found to yield similar results in most cases. \cite{Runeson2023a} Therefore, we use this method of selecting initial conditions due to its ease of implementation.

To measure populations, several different estimators have been proposed in the literature. In this paper we measure the active state, corresponding to an estimator of the form
\begin{equation}
\Theta_n(t) = \begin{cases} 1, & \text{if \space}  P_n(t) > P_m(t) \text{\space} \forall \text{\space} m\neq n \\ 0, & \text{otherwise} \end{cases}
\end{equation}
With this choice of estimator and the initial distribution described above, the ensemble-averaged population transfer from state $i$ to state $n$ is

\begin{equation}
P_{i \rightarrow n}(t) = \langle \Theta_i(0) \Theta_n(t) \rangle
\end{equation}

In the two-state case, Mannouch and Richardson have shown that this prescription overestimates the rate of transfer compared to Landau-Zener theory. It does, however, maintain a physical population estimate with a low number of trajectories. Note, that this approach is different from what is used in both MASH and multistate MASH. Other alternatives can be seen in the SI.

%---

%It's tempting to put some of these details in the SI, but I think it would be harder to read if it's split up than if it's kept together - let me know what you think.}

%\par A population estimator that works with this initial population distribution is given by,
%\begin{equation}
%    \label{eq:population_estimator}
%    \Phi_n = \frac{1}{N} + \alpha_n \Bigl(|c_n|^2 - \frac{1}{N} \Bigl),
%\end{equation}
%\noindent where $\alpha_N = \frac{N-1}{H_N - 1}$ with $H_N =  \sum^N_{n=1} \frac{1}{n}$. Equation \ref{eq:population_estimator} guarantees the previous equation is independent on the basis used and ensures the correct long-time behavior of the populations. One could alternatively measure population as 1 for the active state and 0 for all other states, but this choice would give the wrong Landau-Zener transition probability when used together with the current choice of initial distribution.\cite{Mannouch2023} Furthermore, previous studies have shown that the estimated population, $\Phi_n$, can be modified by applying a weighting function. \cite{Mannouch2023, Runeson2024}

%\YYY{previous statement:This is basis-equivariant and has the correct equilibrium.}  \YYY{Someone has to explain the previous sentence to me!}

\par One potential side-effect of multistate MASH is hopping between states that are not coupled. It is possible to nevertheless accept all hops, regardless of the coupling between the states, which was the approach taken in reference \citenum{Runeson2023a}. However, due to the risk of unphysical behavior, we introduce for all hops an energy criterion, such that $|E_a - E_b| < 0.055$ Hartree ($\approx$ 1.5 eV) to prevent erroneous hops between uncoupled states. Hops that do not meet this criterion are treated on the same footing as reflected hops, with the total velocity of all atoms reflected. This direction of rescaling differs from previous work with MASH and is chosen due to simplicity.
%A new way to scale velocities was also introduced in Ref XXX, this, however, was not employed in our implementation of multi-state MASH so as to avoid the need of calculating non-adiabatic coupling vectors explicitly and maintaining the most favorable qualitity of TSH, the speed of calculations.

%\YYY{I think this sentence would be really good at the beginning of the section where you explain what SH is.} However, the method of hopping is no longer stochastic, multi-state MASH uses a deterministic algorithm by selecting $c_a$ to be the state with the largest electronic wavefunction coefficient. \YYY{Emphasize in the SH part why it is a stochastic method.} 

%  However, whereas TSH assumes a pure excitation into the state of interest, multi-state MASH requires a stochastic sampling over all possible values of $c$ \YYY{The hop is not stochastic but the initial conditions are, what is the benefit of this method then?}, then only initial conditions where the electronic wavefunction coefficient for $c_a$ is greatest are selected. 

\subsection{Ultrafast electron diffraction (UED)\label{sec:IAM}}
\par  We simulate electron diffraction using the independent atom model (IAM), originally devised by Debye.\cite{Debye1930, Bewilongua1932, KirranderWeber2017, SimmermacherCh3Kasra}. This method, extensively used in the analysis of scattering and crystallography experiments, approximates the electron scattering probability, $\big| Z (\bss, \bar{\bR}) \big|^2$, as a coherent sum of tabulated\cite{IntTabCryVolC} form factors, $f_N^{\mathrm{e}} (s)$,
\begin{align} \label{eq:IAM}
\big| Z (\bss, \bar{\bR}) \big|^2 = \sum_{A}^{N_{\mathrm{at}}} \sum_{B}^{N_{\mathrm{at}}}\ f_A^{\mathrm{e}} (s)\ f_B^{\mathrm{e}} (s)\ e^{\mathrm{i} \bss\cdot\bR_{AB}}\text{,}
\end{align}
where $\bss$ is the momentum transfer vector, $\bar{\bR}$ the molecular geometry, $\bR_{AB}$ the distance vector between atoms $A$ and $B$, and $N_{\mathrm{at}}$ the number of atoms in the molecule. Note that $\big| Z (\bss, \bar{\bR}) \big|^2$ is given in units of the Rutherford cross-section (see \textit{e.g.}\ reference \citenum{SimmermacherCh3Kasra} for further details). The electron scattering atomic form factors, $f_A^{\mathrm{e}} (s)$, are defined as, 
\begin{align} \label{eq:formf}
f_A^{\mathrm{e}} (s) = f_A^{\mathrm{x}} (s) - Z_{A}\text{,}  
\end{align}
where $f_A^{\mathrm{x}} (s)$ are the tabulated form factors for x-rays and $Z_{A}$ the nuclear charges. 

We calculate the rotationally averaged signal accounting for all possible molecular orientations in a thermal ensemble. This yields the isotropic elastic electron scattering probability, $\mathcal{S}_{0} (s, \bar{\bR})$, which takes the form, 
\begin{align} \label{eq:IAM-RA}
\mathcal{S}_{0} (s, \bar{\bR}) = \sum_{A}^{N_{\mathrm{at}}} \sum_{B}^{N_{\mathrm{at}}}\ f_A^{\mathrm{e}} (s)\ f_B^{\mathrm{e}} (s)\ j_{0} (sR_{AB})\text{,}
\end{align}
where $s=|\bss|$ is the norm of the momentum transfer vector and $R_{AB}=|\bR_{AB}|$ is the distance between atoms $A$ and $B$, and where we have introduced the zeroth-order spherical Bessel function $j_{0} (sR_{AB})$ defined as, 
\begin{align} \label{eq:Bessel}
j_{0} (sR_{AB})=\frac{\sin{(sR_{AB})}}{sR_{AB}}\text{.}
\end{align}

In the experiment, not only the elastic component of electron scattering is measured but also the inelastic component. While the IAM cannot describe individual inelastic transitions, the total inelastic component can be approximated by an incoherent sum of {\emph{atomic}} inelastic scattering functions $S_A(s)$. This yields the total isotropic scattering, $\mathcal{S}_{t} (s, \bar{\bR})$, as the sum of the elastic and inelastic contributions,
\begin{align} \label{eq:IAM-RA-tot}
\mathcal{S}_{t} (s, \bar{\bR}) = \mathcal{S}_{0} (s, \bar{\bR})+\sum_{A}^{N_{\mathrm{at}}} S_A(s) \text{.}
\end{align}

We note that further improvements in the scattering signal are straightforward. For instance, alignment effects resulting from the linear polarization of the pump laser can be accounted for\cite{SimmermacherPRA2020} and corrections to the form factors due to relativistic effects can be made,\cite{ELSEPA2005} however for the latter we note that this effect is comparatively minor for the electron energies at the SLAC MeV-UED source. It is also possible to calculate the scattering cross-sections from {\it{ab initio}} electronic wavefunctions, which makes it possible to account accurately for the effects on scattering from chemical bonding, electron correlation, and change in the electronic state due to excitation by the pump pulse. \cite{Carrascosa2019total,Zotev2020,Moreno2022}
%of bonding electrons and inelastic transitions.\cite{Carrascosa2019total,Zotev2020,Moreno2022}

\section{\label{Methods}Computional details}

%\YYY{(Adam: Should we explain somewhere why we decided that we do not need to include the triplet states?)}

\par A ground-state minimum energy structure of cyclobutanone was obtained using state-averaged CASSCF,\cite{Roos1980a} SA(3)-CASSCF(12,12)/aug-cc-pVDZ, \textit{i.e.}\ state-averaging over the three lowest singlet states, an active space with 12 electrons in 12 orbitals with the aug-cc-pVDZ Cartesian basis.\cite{Woon1993a} Further details are given in the supplementary information. All electronic structure calculations were done in OpenMolcas version v23.02-10-gb7266214b.\cite{Aquilante2020, LiManni2023}  The ground-state minimum energy structure was subjected to a frequency calculation yielding no imaginary frequencies, confirming the structure to be a true minimum. Further excited state structures were optimized using SA(3)-CASSCF(12,12) to obtain the S$_2$ minimum, two S$_2$/S$_1$ minimum energy conical intersections (MECIs), and two S$_1$/S$_0$ MECIs. Selected structures were subjected to linear interpolation in internal coordinates (LIIC) to yield a progression of intermediate structures. Potential energy curves along the LIICs were calculated using SA(3)-CASSCF(12,12) and extended multistate complete active space second-order perturbation theory, XMS-CASPT2, with the same (12,12) active space as for the CASSCF calculations (see supplementary information).\cite{Shiozaki2011a} For stability, an imaginary shift of 0.5 a.u.\ was applied for all XMS-CASPT2 calculations.\cite{Forsberg1997a} 

For further comparison, the potential energy curves were also calculated using a modified version of adaptive configuration interaction (ACI) and Monte-Carlo configuration interaction (MCCI). \cite{Schriber2016, Schriber2017a, Prentice2023} The state-averaged MCCI calculations were performed using MCCI V4\cite{Tong2000, Coe2013} with the SA error-controlled calculations performed using a locally modified development version of GeneralSCI 1.0.\cite{Prentice2023b} Further details are given in the supplementary information.% \YYY{(ADAM: I would take all the text from Andrew below and put it, without modification, in the SI.)}

\par The calculated ground state frequencies were used to generate a Wigner distribution from which 1000 geometries were sampled. At each geometry, excitation energies and oscillator strengths were calculated using SA(3)-CASSCF(12,12). The photoabsorption cross-section was constructed using the nuclear ensemble approach (NEA), convoluted by Lorentzian functions with a phenomenological broadening parameter of 0.05 eV.\cite{Crespo-Otero2012e} Initial conditions for the multistate MASH were sampled from this Wigner distribution using an implicit laser pulse of width 0.2 eV, centered at 6.2 eV. All trajectories were initiated from the bright S$_2$ (n-3s Rydberg) state using multistate MASH (see section \ref{sec:MASH} for more details) \cite{Runeson2023a} implemented in a development version of SHARC.\cite{Mai2018a} The electronic structure calculations were carried out using SA(3)-CASSCF(12,12)/aug-cc-pVDZ with all three electronic states available in dynamics. The trajectories were propagated using a 0.5 fs time step, with 25 substeps, for 300 fs using a local diabatization scheme. For trajectories in which the total energy changes by more than 0.25 eV, the propagation is stopped and only the time points with the correct total energy are used in the analysis. Additionally, we note that a subset of trajectories are 'trapped'. This occurs when the trajectories enter a region where a reflected hop happens but the reflected hop fails to remove the trajectory from the 'forbidden' region. Such trapped trajectories are only included in the analysis until the point that they get trapped. For a summary of the trajectories removed from the analysis, see the supplementary information. Populations of the ensemble of trajectories are estimated using the multistate Populations were estimated according to Section \ref{sec:MASH}. Finally, rotationally averaged total scattering diffraction patterns were calculated using the geometries from the multistate MASH trajectories and the independent atom model (see Section \ref{sec:IAM}) for the initial 200 fs of dynamics.

\section{\label{Results}Results and discussion}
\subsection{Excited states of cyclobutanone}

\par The character of the two lowest-lying S$_1$ and S$_2$ electronic states of cyclobutanone has previously been identified as n$\pi^*$ and n-3s, respectively.\cite{Kuhlman2012, Drury-Lessard1978} The next set of excited states are the higher-lying 3p Rydberg states, which appear a full 0.7 eV above the n-3s Rydberg state.\cite{OToole1991} Triplet states have been shown to be important for the photodynamics upon excitation with long wavelengths.\cite{Kao2020} However, due to the short excited state lifetime of cyclobutanone following excitation into the S$_2$ state, triplet states are less likely to make a significant contribution to the current photodynamics. For these reasons, we have chosen to focus our attention on an accurate description of the two lowest-lying singlet states S$_1$ and S$_2$. 

\par A summary of the low-lying excited singlet states available to cyclobutanone at the equilibrium geometry of cyclobutanone is shown in Table \ref{tab:excitation_energies}. Both SA(3)-CASSCF(12,12) and XMS-CASPT2(12,12) yield an S$_1$ state with n$\pi^*$ character at the Franck-Condon (FC) geometry and an S$_2$ state with n-3s Rydberg character. Furthermore, the excitation energies for SA(3)-CASSCF(12,12) are in excellent agreement with XMS-CASPT2(12,12), with SA(3)-CASSCF(12,12) energies showing a constant difference of 0.04 eV for both excited states.  Excitation energies for SA(3)-ACI and SA(3)-MCCI can also be found in Table \ref{tab:excitation_energies}. These methods give a good description of the low-lying singlet states of cyclobutanone with excitation energies deviating by $\sim$0.05 eV for the valence n$\pi^*$ state and $\sim$0.1 eV for the n-3s Rydberg state compared to the XMS-CASPT2(12,12) benchmark. In the ACI simulation, a large orbital window, corresponding to a CAS-CI(12,36), has also been tested. The large orbital window gives results analogous to the ones obtained with CASSCF(12,12), providing further support for this choice of active space.   %deviating by 0.04 eV for both states of interest. 

%The photochemical reactivity of cyclobutanone has been previously studied after photoexcitation into S$_1$ (n$\pi^*$) finding a ring-opening mechanism to be the dominant decay pathway. Previous studies have also attributed the second singlet state to have n-3s Rydberg character.

\begin{table}
\caption{\label{tab:excitation_energies} Table of excitation energies in eV for the lowest two excited electronic states of cyclobutanone at the S$_0$ minimum energy geometry optimized using SA(3)-CASSCF(12,12)/aug-cc-pVDZ. State characters are given in parentheses along with excitation energies.}
\begin{ruledtabular}
\begin{tabular}{ccc}
Method & S$_1$ & S$_2$\\
\hline
SA(3)-CASSCF(12,12) & 4.44 (n$\pi^*$) & 6.24 (n-3s)\\
XMS-CASPT2(12,12) & 4.40 (n$\pi^*$) & 6.20 (n-3s)\\
SA(3)-ACI & 4.43 (n$\pi^*$) & 6.31 (n-3s) \\
SA(3)-MCCI & 4.45 (n$\pi^*$) & 6.30 (n-3s) \\
\end{tabular}
\end{ruledtabular}
\end{table}

\par
\begin{figure}
    \centering
    \includegraphics[width=0.49\textwidth]{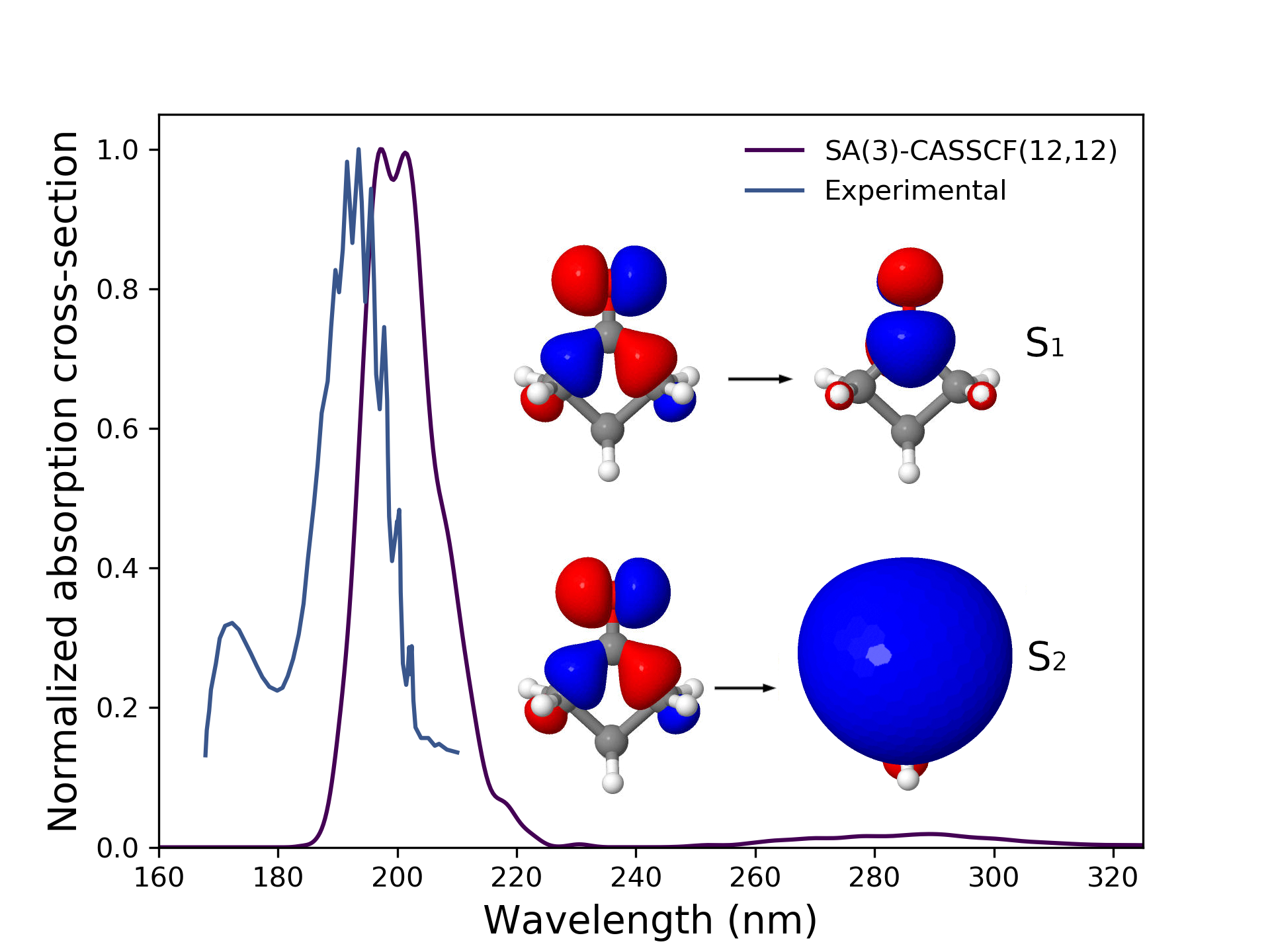}
    \caption{Photoabsoprtion cross-section of gas-phase cyclobutanone calculated using the SA(3)-CASSCF(12,12)/aug-cc-pVDZ, the NEA and Wigner sampling of 1000 geometries (purple line). Experimental data is reproduced from reference \citenum{Udvarhazi1965} (blue line). Orbitals for the S$_1$ and S$_2$ states are shown as an inset.}
    \label{fig:spectra}
\end{figure}

 A theoretical photoabsorption cross-section of cyclobutanone calculated using SA(3)-CASSCF(12,12) is shown in Fig.\ \ref{fig:spectra}, allowing for a direct comparison with experimental absorption spectra. Good agreement can be observed between the two spectra, further validating our electronic structure method along with confirming the assignment of states using the insets in Fig.\ \ref{fig:spectra}.

%Additionally, the experimentally determined absorption spectrum is plotted confirming the assignment of these states and showing good agreement for our SA(3)-CASSCF(12/12) photoabsorption cross-section.

\subsection{\label{sec: pathways}Photochemical reaction pathways and benchmarking}
\par
\begin{figure}
    \centering
    \includegraphics[width=0.49\textwidth]{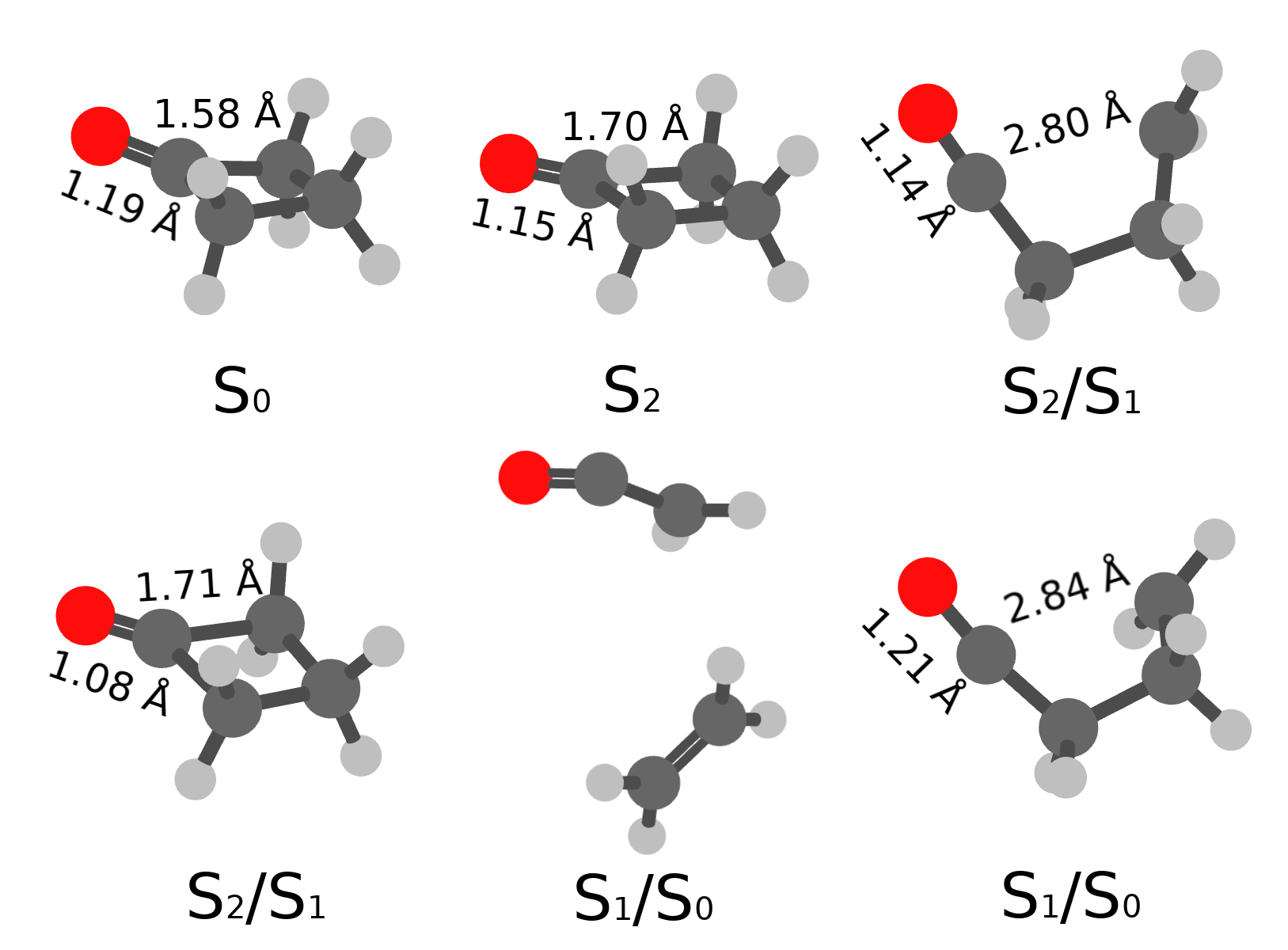}
    \caption{Six critical geometries are shown for gas-phase cyclobutanone. The S$_0$ minimum (upper left), S$_2$ minimum (upper center), a dissociative S$_2$/S$_1$ MECI (upper right), another S$_2$/S$_1$ MECI (lower left), a ring breaking S$_1$/S$_0$ MECI (lower center), and a ring-opening S$_1$/S$_0$ (lower right). All structures shown are optimized using SA(3)-CASSCF(12,12)/aug-cc-pVDZ.}
    \label{fig:structures}
\end{figure}

  After the photophysics of cyclobutanone has been studied in the FC region, the different deactivation pathways available upon photoexcitation can be characterized. A series of potentially important optimized geometries for these pathways are shown in Fig.\ \ref{fig:structures}, optimized using SA(3)-CASSCF(12,12)/aug-cc-pVDZ. The S$_0$ minimum energy geometry (S$_0$, upper left) can be characterized by a single carbon atom breaking the planar symmetry of the molecule, along with a \ce{C=O} bond length of 1.19 \AA \space and an $\mathrm{\alpha}$-CC bond of 1.58 \AA. When the molecule is excited by a 200 nm pulse, the S$_2$ excited electronic state is populated at the FC region and can potentially relax into the S$_2$ minimum. The S$_2$ minimum energy geometry (S$_2$, upper center) displays subtle changes in bond lengths, along with a planarization of the carbon ring. Two S$_2$/S$_1$ minimum energy conical intersections (MECIs) were successfully optimized, with the S$_2$/S$_1$ MECI (S$_2$/S$_1$, lower left) already reported previously.\cite{Kuhlman2012a} 
This MECI displays a rather severe compression of the \ce{C=O} bond. The second S$_2$/S$_1$ MECI (S$_2$/S$_1$, upper right) exhibits a dissociated $\mathrm{\alpha}$-CC bond alongside a compressed \ce{C=O} bond. Two further MECIs are included in Fig.\ \ref{fig:structures}, corresponding to MECIs between the ground and first excited state. Both these S$_1$/S$_0$ MECIs have been reported previously by Liu \emph{et al.}.\cite{Liu2016} The first of these MECIs (S$_1$/S$_0$, lower center), termed CI-3 in reference \citenum{Liu2016}, breaks the ring structure to produce \ce{CH_2=CH_2} and \ce{O=CCH_2}. The second S$_1$/S$_0$ MECI (S$_1$/S$_0$, lower right), termed CI-1 in reference \citenum{Liu2016}, is closely related to the second S$_2$/S$_1$ MECI (S$_2$/S$_1$, upper right), with an dissociated $\mathrm{\alpha}$-CC bond. However, in contrast to the S$_2$/S$_1$ MECI, the S$_1$/S$_0$ MECI contains a slightly stretched \ce{C=O} bond relative to the S$_0$ geometry. It is obvious from this static perspective of excited state geometries, that cyclobutanone possesses a rich photochemistry. However, the relative importance of each of these structures is still unknown, as are the final products of each conical intersection (CI).

\par To gain a better understanding of the PESs of cyclobutanone and to benchmark SA(3)-CASSCF(12,12) away from the FC region, an LIIC was done on some of the key geometries shown in Fig.\ \ref{fig:structures}. Four structures were selected: the S$_0$ minimum (Fig.\ \ref{fig:structures}, S$_0$, upper left),  S$_2$ minimum (Fig.\ \ref{fig:structures}, S$_2$, upper center), the dissociative S$_2$/S$_1$ MECI (Fig.\ \ref{fig:structures}, S$_2$/S$_1$, upper right) and finally the $\mathrm{\alpha}$-CC bond dissociation S$_1$/S$_0$ MECI (Fig.\ \ref{fig:structures}, S$_1$/S$_0$, upper right).

\begin{figure}
    \centering
    \includegraphics[width=0.49\textwidth]{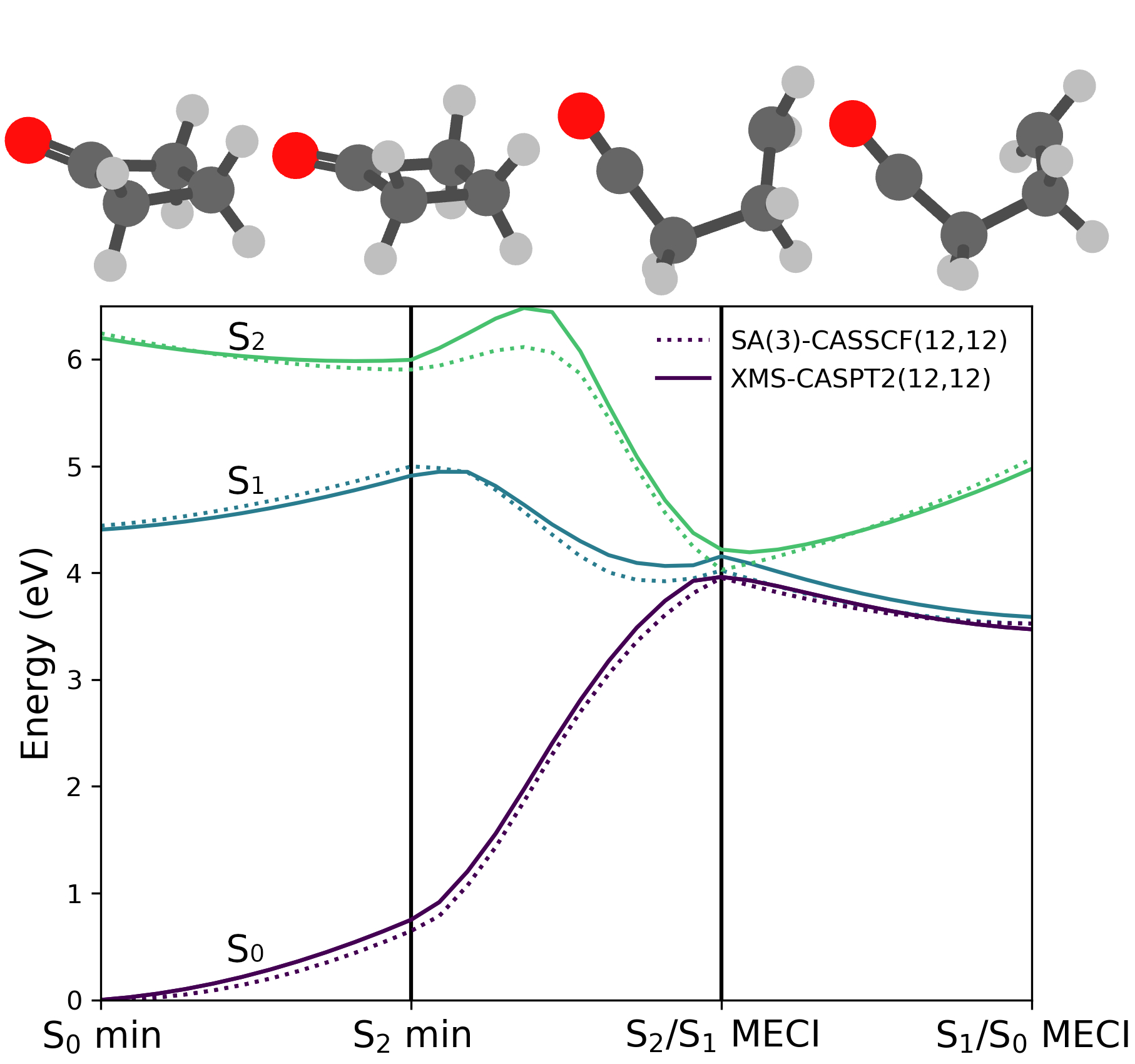}
    \caption{A linear interpolation in internal coordinates from the S$_0$ minimum to S$_2$ minimum to the dissociative S$_2$/S$_1$ MECI ending with the ring opening S$_1$/S$_0$ MECI. Inset structures correspond to the optimized structures used along the LIIC coordinate. Excitation energies for the three lowest-lying singlet states are given using both SA(3)-CASSSCF(12,12)/aug-cc-pVDZ (dotted line) and XMS-CASPT2(12,12)/aug-cc-pVDZ (solid line).}
    \label{fig:LIIC}
\end{figure}

\par Upon photoexcitation to the S$_2$ state, cyclobutanone in the S$_0$ minimum structure can relax to the closeby S$_2$ minimum via a planarization of the carbon ring (see green line in panel 1 of Fig.\ \ref{fig:LIIC}). Here, cyclobutanone undergoes a dissociation of an $\mathrm\alpha$-CC bond resulting in a barrier height of 0.21 eV for SA(3)-CASSCF(12,12), somewhat below the 0.49 eV predicted by XMS-CASPT(12,12), with the barrier heights given as energy differences relative the S$_2$ minimum. Note that absolute values for transition state barriers in LIICs should be treated with care as they are known to be overestimated. Regardless, SA(3)-CASSCF(12,12) is likely to estimate a faster rate of decay from S$_2$ to S$_1$ as a result of this smaller reaction barrier. A steep drop in energy is observed for both methods after the transition barrier leading down to the S$_2$/S$_1$ MECI (see Fig.\ \ref{fig:structures}, upper right and inset of Fig.\ \ref{fig:LIIC}). This S$_2$/S$_1$ MECI is peaked and has a single pathway according to the criteria set out in reference \citenum{Fdez.Galvan2016}. This allows for efficient funneling onto the S$_1$ state (see blue line in Fig.\ \ref{fig:LIIC}) where a barrier-less decay is observed to the structurally similar S$_1$/S$_0$ MECI (see Fig.\ \ref{fig:structures}) where the S$_1$/S$_0$ MECI is observed with a peaked bifurcating branching space. 

\par From Fig.\ \ref{fig:LIIC} it can be deduced that there is excellent agreement between SA(3)-CASSCF(12,12) and XMS-CASPT2(12,12), with slight differences observed for the approximate transition barrier height. Nevertheless, overall the SA(3)-CASSCF(12,12) results offer an excellent comprise between speed and accuracy for the {\it{on-the-fly}} nonadiabatic dynamics of gas-phase cyclobutanone. Additional LIICs for the same geometries can be found in SI (Figures 3 and 4) for both SA(3)-ACI and SA(3)-MCCI. These further support our assessment. We also note that this indicates that good agreement can be obtained with methods that potentially offer a more black-box approach to the calculation of multireference wavefunctions than the CAS family of methods.

%Where XMS-CASPT2(12/12) maintains a splitting between S$_2$ and S$_1$ of 0.063 eV indicating XMS-CASPT2(12/12) is likely close to a conical intersection

\subsection{Dynamics}

\par A total of $229$ trajectories were initiated on the S$_2$ state. Trajectories were removed from the statistics if either a change in total energy of $0.25$ eV or more was observed or if consecutive reflected hops indicated that the trajectory was trapped in a 'forbidden' region of the PES. This resulted in a significant number of discarded trajectories. For a full summary of discarded trajectories see the SI (Figure 2). To briefly summarize, after $50$ fs the number of trajectories had dropped to $146$, and by $200$ fs only $18$ trajectories of the initial $229$ trajectories had survived, while at $300$ fs this was down to $10$ trajectories. At $300$ fs a total of $71$ trajectories had been discarded due to consecutive reflected hops, while $148$ trajectories had been discarded due to issues with the energy conservation. We acknowledge that this could potentially cause artifacts in the simulations, especially if certain processes are systematically removed or enhanced by the removal of trajectories.% In addition, it means our simulations will be far from converging at later time points. 

\begin{figure}[h!]
    \centering
    \includegraphics[width=0.49\textwidth]{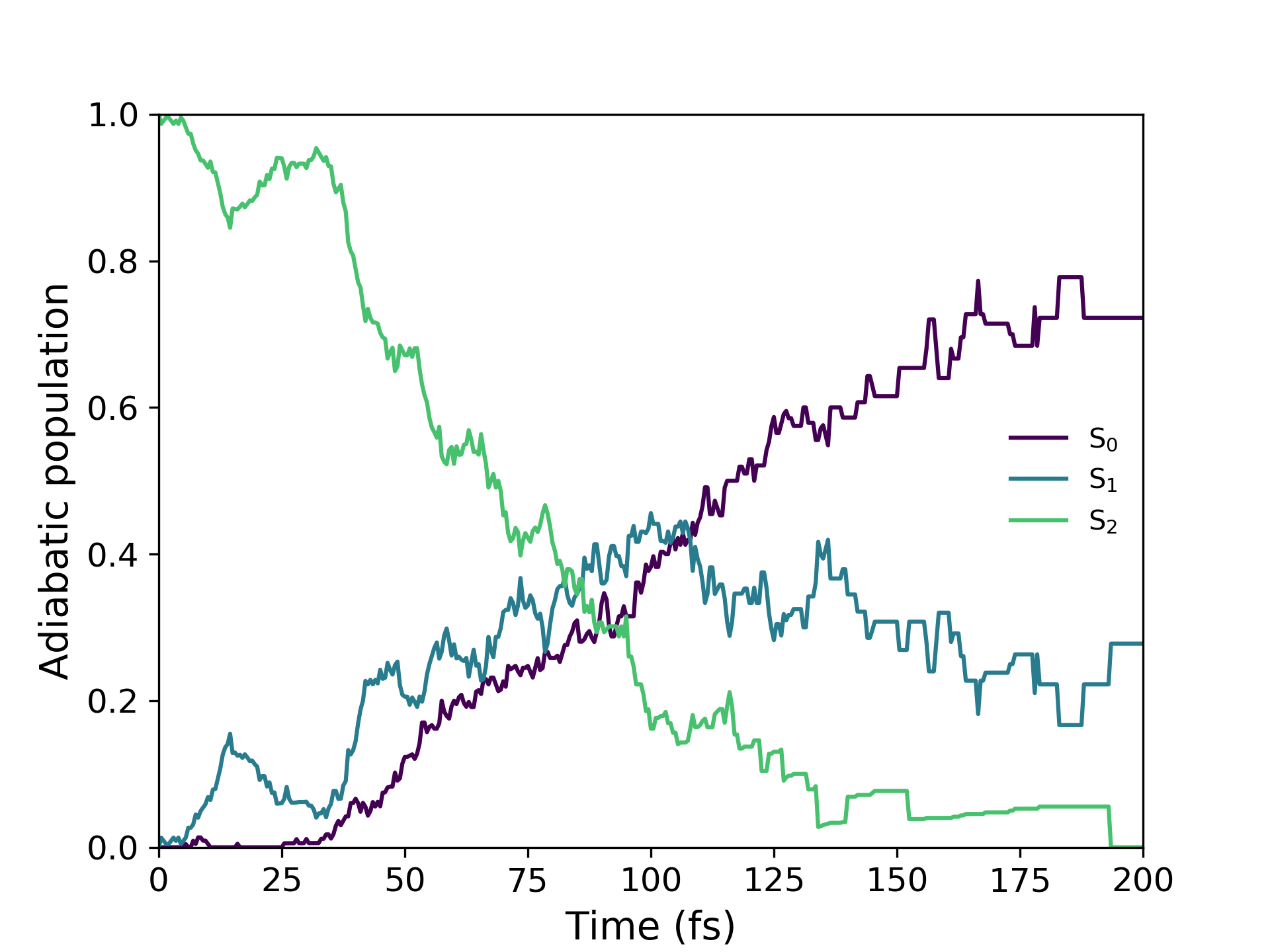}
    \caption{Adiabatic populations for the three lowest-lying singlet states, S$_{0-2}$, given by the multistate MASH population estimator (Section \ref{sec:MASH}) given as a function of time and taking all trajectories into account.}
    \label{fig:Population}
\end{figure}
% according to the multistate MASH population estimator (given in Equation \ref{eq:population_estimator})

\par The adiabatic populations for the three electronic states S$_{0-2}$ used in our simulations are given in Fig.\ \ref{fig:Population} for the initial $200$ fs of the dynamics. A rapid decay of the S$_2$ state (green line) can be observed after $\sim$8 fs. This decay is correlated with a population increase on S$_1$ (teal line), indicating that fast initial dynamics on S$_2$ rapidly reaches an S$_2$/S$_1$ CI. For the early time dynamics ($t<30$ fs), the ground-state can be seen to be inaccessible from the S$_2$ state (see panel 1 in Fig.\ \ref{fig:LIIC}). After $30$ fs, there is a steady increase in the ground-state S$_0$ (purple line) population, showing passage through the S$_1$/S$_0$ CI and therefore formation of the photoproducts linked to each CI (Fig.\ \ref{fig:structures}). After $100$ fs, we observe a decreased rate of population transfer from S$_2$ to S$_1$, resulting in depletion of S$_1$ and an overall decrease in the population of the two excited states matched by the continued increase in the population of the ground state. After $\sim175$ fs, the population transfer is largely over, although errors in this region will be high due to the small number of trajectories. The populations observed in Fig.\ \ref{fig:Population} indicate ultrafast non-radiative deactivation pathways for cyclobutanone, there is then likely the possibility of further non-equilibrium ground-state chemistry as a result of very hot molecules in the ground state following the decay dynamics.

\par
\begin{figure}[h!]
    \centering
    \includegraphics[width=0.49\textwidth]{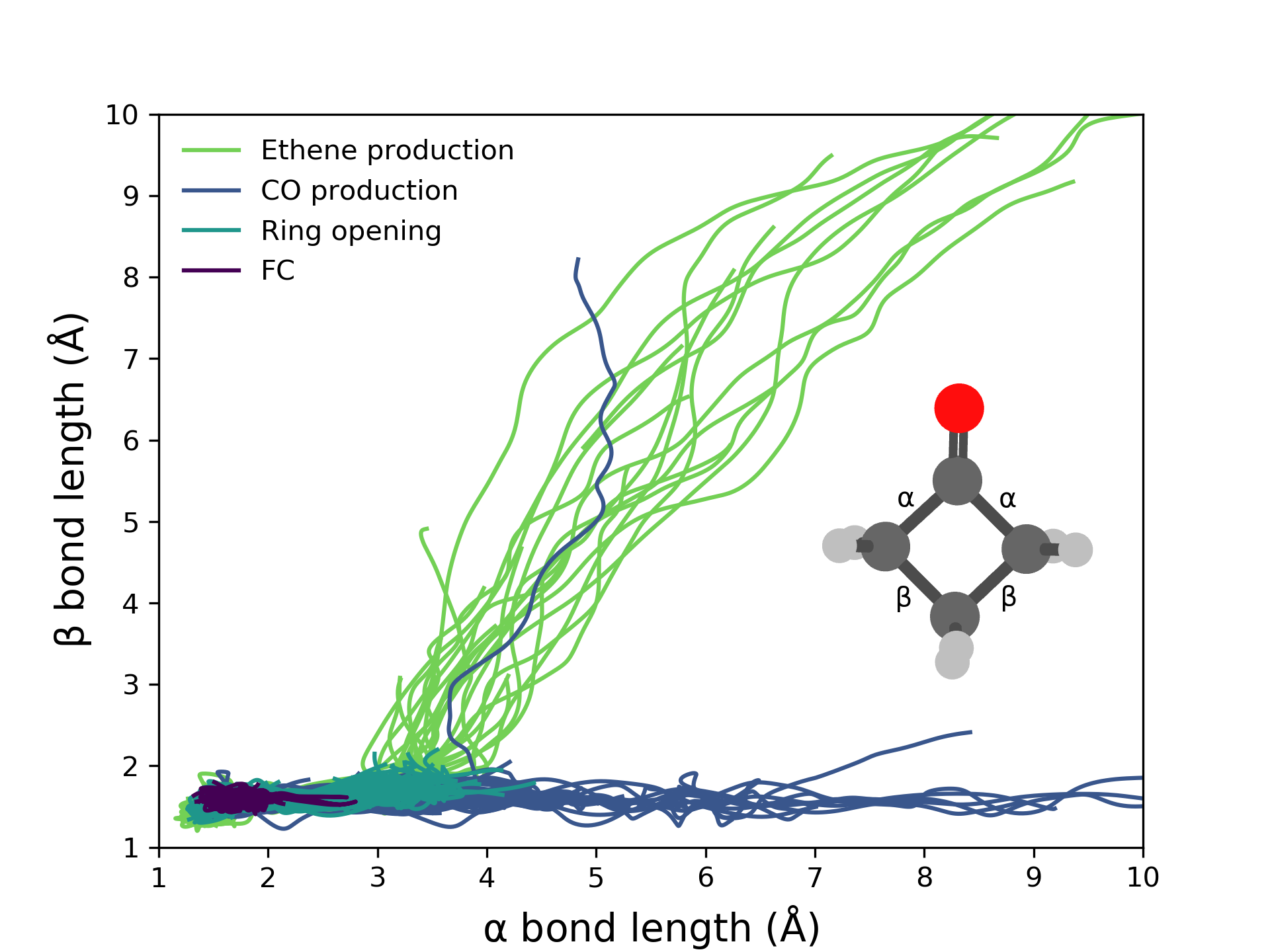}
    \caption{A 2D projection of all \ce{CC} bond lengths across all trajectories. The \ce{CC} bonds are partitioned according to the labeling on the inset. Each trajectory is clustered based on the final geometry of the trajectory, resulting in four channels: \emph{i)} FC, \emph{ii)} ring opening, \emph{iii)} CO production and \emph{iv)} ethene production. See the main text for more details.  }
    \label{fig:bonds}
\end{figure}

 As shown previously, the deactivation pathways available to cyclobutanone imply the cleavage of one or two bonds in the ring structure. To identify these different pathways, the bond lengths for all multistate MASH trajectories are plotted, separated into $\mathrm{\alpha}$ and $\mathrm{\beta}$-CC bonds according to the inset in Fig.\ \ref{fig:bonds}. Four distinct outcomes are observed:
\begin{itemize}
    \item[\emph{(i)}] There is no bond breaking (purple), and the molecule remains in the FC region. These are characterized by short $\mathrm{\alpha}$ and $\mathrm{\beta}$-CC bond lengths ($R<2.2$ \AA). It is important to note that the trajectories included in this group do not progress over 90 fs. 
    \item[\emph{(ii)}] There exists a single $\mathrm{\alpha}$-CC bond breaking (teal). This pathway gives rise to a ring-opened structure {\it{i.e.}} \ce{CH_2CH_2CH_2CO}. This is indicated by an elongation of an $\mathrm{\alpha}$-CC bond, but due to the carbon backbone still being intact, the distance between the two carbon atoms is limited to $R\sim4.5$ \AA. This pathway is a result of passing through the S$_1$/S$_0$ CI (upper right in Fig.\ \ref{fig:structures}).
    \item[\emph{(iii)}] Two $\mathrm{\alpha}$-CC bonds break to liberate \ce{CO}, where there is a large increase in $\mathrm{\alpha}$-CC bonds to values above 4.5 \AA \space (blue). Attempts to optimize an MECI using SA(3)-CASSCF(12,12) involving \ce{CO} dissociation were unsuccessful, however, trajectories were observed with \ce{CO} dissociation after having decayed to the ground state via the $\mathrm{\alpha}$-CC dissociation S$_1$/S$_0$ CI (lower right, Fig.\ \ref{fig:structures}). This means the \ce{CO} dissociation takes place on the ground state rapidly after passing through a CI. It is worth noting for option \emph{iii)} a single trajectory also displays ground state dissociation of \ce{CH_2CH_2CH_2} into \ce{CH_2CH_2} and \ce{CH_2}, which can be seen in Fig.\ \ref{fig:bonds} as a blue line displaying increases in both $\mathrm{\alpha}$ and $\mathrm{\beta}$-CC bonds.
    \item[\emph{(iv)}] An $\mathrm{\alpha}$-CC bond and $\mathrm{\beta}$-CC bond breaking (pale green) showing the stepwise ring-breaking mechanism where ethene and ketene are produced. Here, an $\mathrm{\alpha}$-CC bond breaks first, followed by a $\mathrm{\beta}$-CC bond breaking. This can be linked to the S$_1$/S$_0$ MECI in Fig.\ \ref{fig:structures} (S$_1$/S$_0$, lower center).
\end{itemize}

%\par To identify all deactivation pathways available to cyclobutanone in our simulations, the bond lengths for all trajectories are plotted and grouped according to the inset in Figure \ref{fig:bonds}. Four outcomes can be observed: \emph{i)} cyclobutanone in the FC geometry, \emph{ii)} a single $\mathrm{\alpha}$-CC bond breaking, \emph{iii)} two $\mathrm{\alpha}$-CC bonds breaking to liberate \ce{CO} and \emph{iv)} an $\mathrm{\alpha}$-CC bond and $\mathrm{\beta}$-CC bond breaking to produce ethene and ketene.

\subsection{UED}

\begin{figure}
    \centering
    \includegraphics[width=0.51\textwidth]{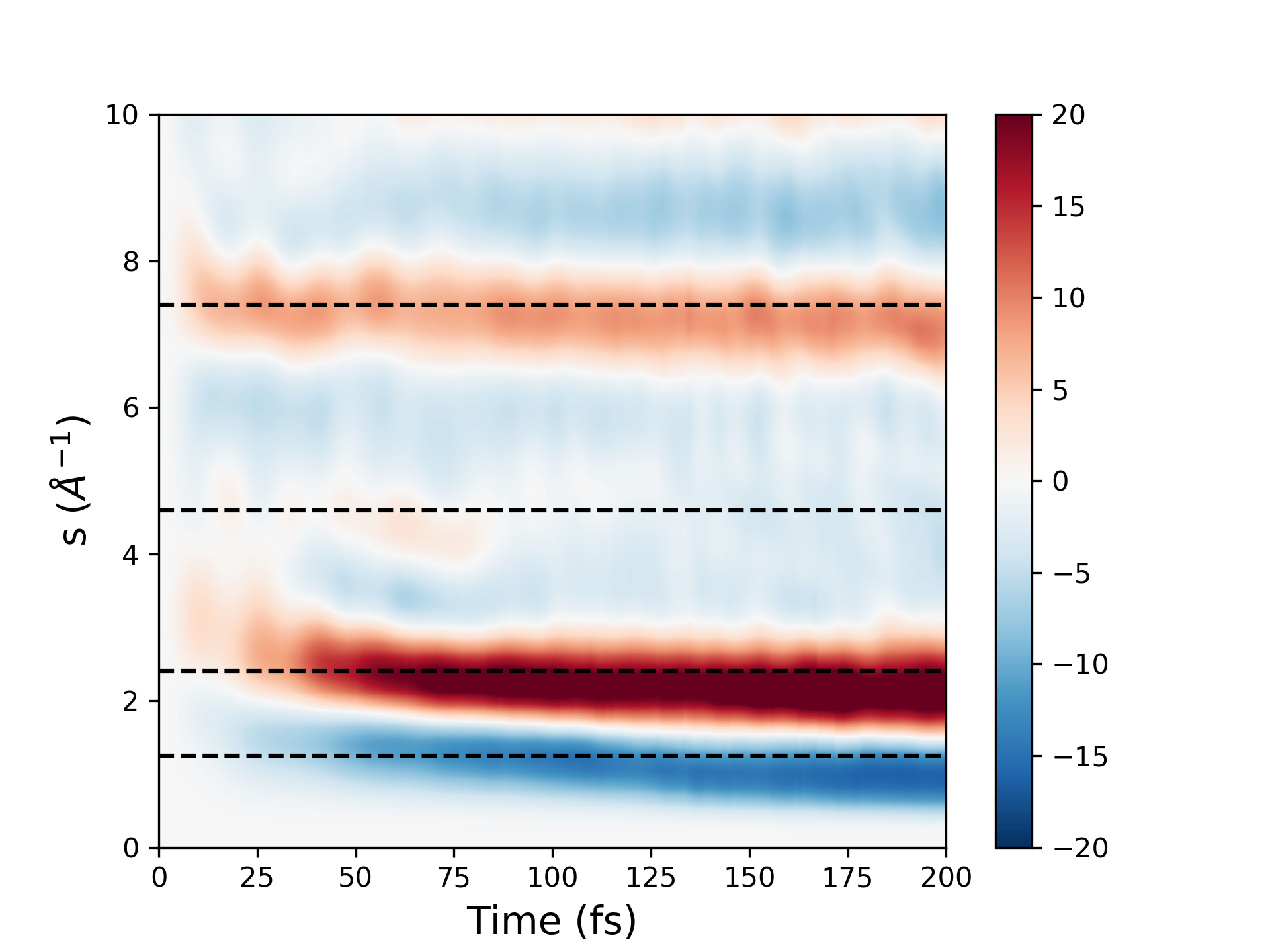}
    \caption{Gas-phase UED pattern shown with percent difference as a function of time for cyclobutanone. All active trajectories are taken into account via the process described in section \ref{Methods}. Dashed horizontal lines represent the key features of the UED pattern.} 
    \label{fig:ued_full}
\end{figure}

%Intro
The nonadiabatic dynamics of cyclobutanone presented in the previous section has served as the basis for calculating the rotationally averaged total electron scattering signal as a function of time, using the independent atom model (IAM) as described in the theory section. To correctly account for the time evolution of the molecular geometries we use a percent difference in the form, 
%\par After simulating the nonadiabatic dynamics of cyclobutanone we can go onto calculate the rotationally averaged gas-phase UED pattern for cyclobutanone shown in Figure \ref{fig:ued_full}, plotted as percentage difference,

\begin{equation}
    \%S_t(s,t) = 100\times\frac{S_t(s,t) - S_t(s,0)}{S_t(s,0)},
\end{equation}

\noindent where $S_t(s,t)$ is the total electron scattering probability at time $t$ and $S_t(s,0)$ is the total electron scattering probability at $t=0$ ({\it{i.e.}} before the pump). In Fig.\ \ref{fig:ued_full}, the time-resolved UED pattern for cyclobutanone is shown as a function of the amplitude of the momentum transfer vector $s$. Due to the rapid loss of trajectories during the simulations, discussed earlier, we only show the UED signal for the first 200 fs.

%\noindent where $S_t(s,t)$ is the total electron scattering probability at time $t$ and $S_t(s,0)$ is the total electron scattering probability at $t=0$. It is important to note that all probabilities are rotationally averaged and presented in Rutherford cross-section units for simplicity.  Due to the steep loss of trajectories during our simulations we only simulate the first 200 fs for our UED, even with this truncation, the statistics at later time points are likely far from converging. All data required for Figure \ref{fig:ued_full} is supplied in XXX to allow for better comparisons to experimental results.  

\begin{figure*}
    \centering
    \includegraphics[width=0.99\textwidth]{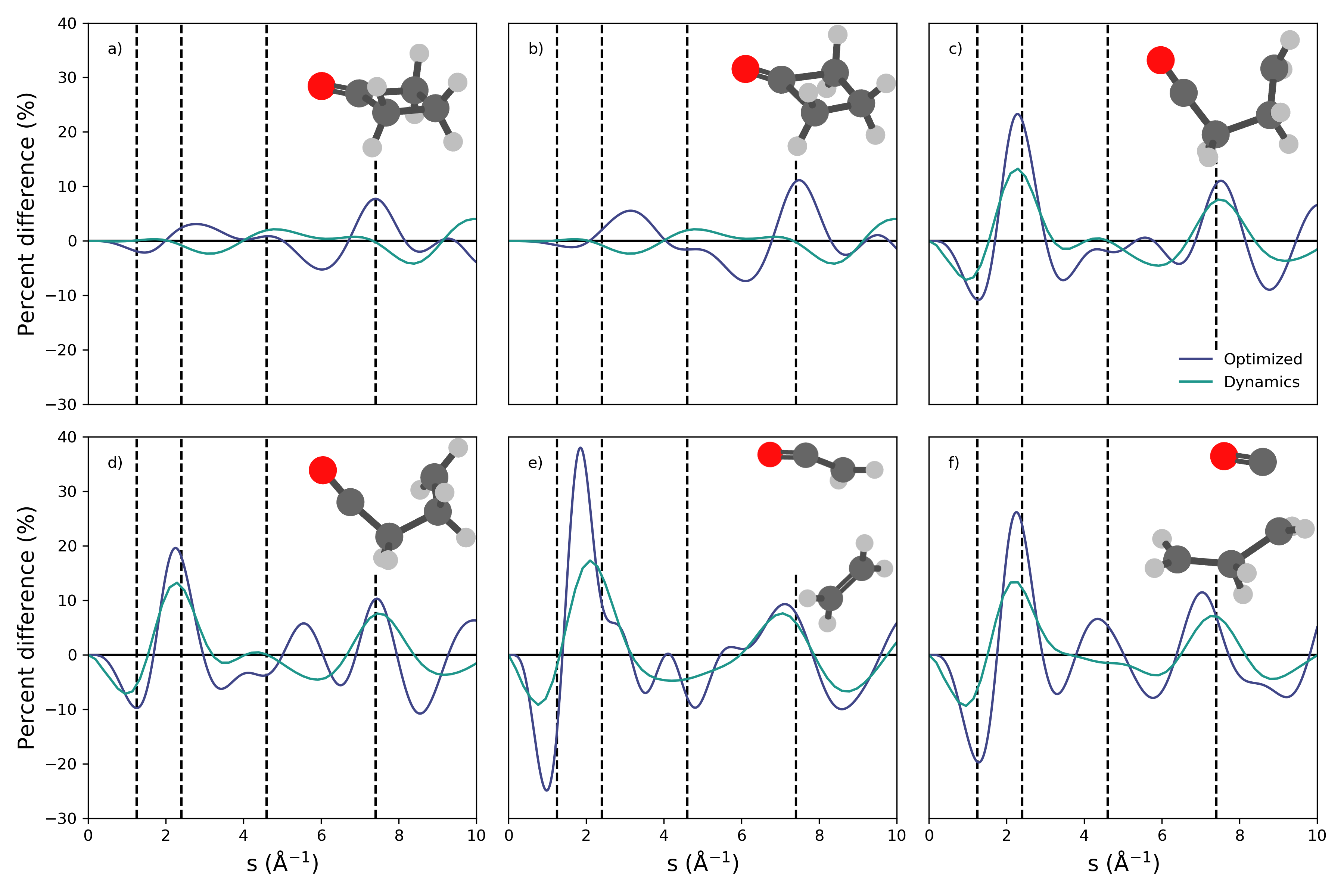}

    \caption{Six static UED percent difference signals for key optimized geometries of gas-phase cyclobutanone and the static signal from an example structure obtained from our dynamics simulations showing \ce{CO} dissociation (blue) and an average of the trajectories where the considered structure is the final product (green). Insets show the structure used to calculate the static signal: a) S$_2$ minimum, b) S$_2$/S$_1$ compression, c) S$_2$/S$_1$ $\mathrm{\alpha}$ cleavage, d) S$_1$/S$_0$ $\mathrm{\alpha}$ cleavage, e) S$_2$/S$_1$ ethene production and finally f) \ce{CO} production. Vertical dashed lines match the horizontal dashed lines shown in Figure \ref{fig:ued_full}, showing key features in the overall UED.}
    \label{fig:static_signal}
\end{figure*}

%Summarise main features in UED

\par
\begin{figure}[h!]
    \centering
    \includegraphics[width=0.49\textwidth]{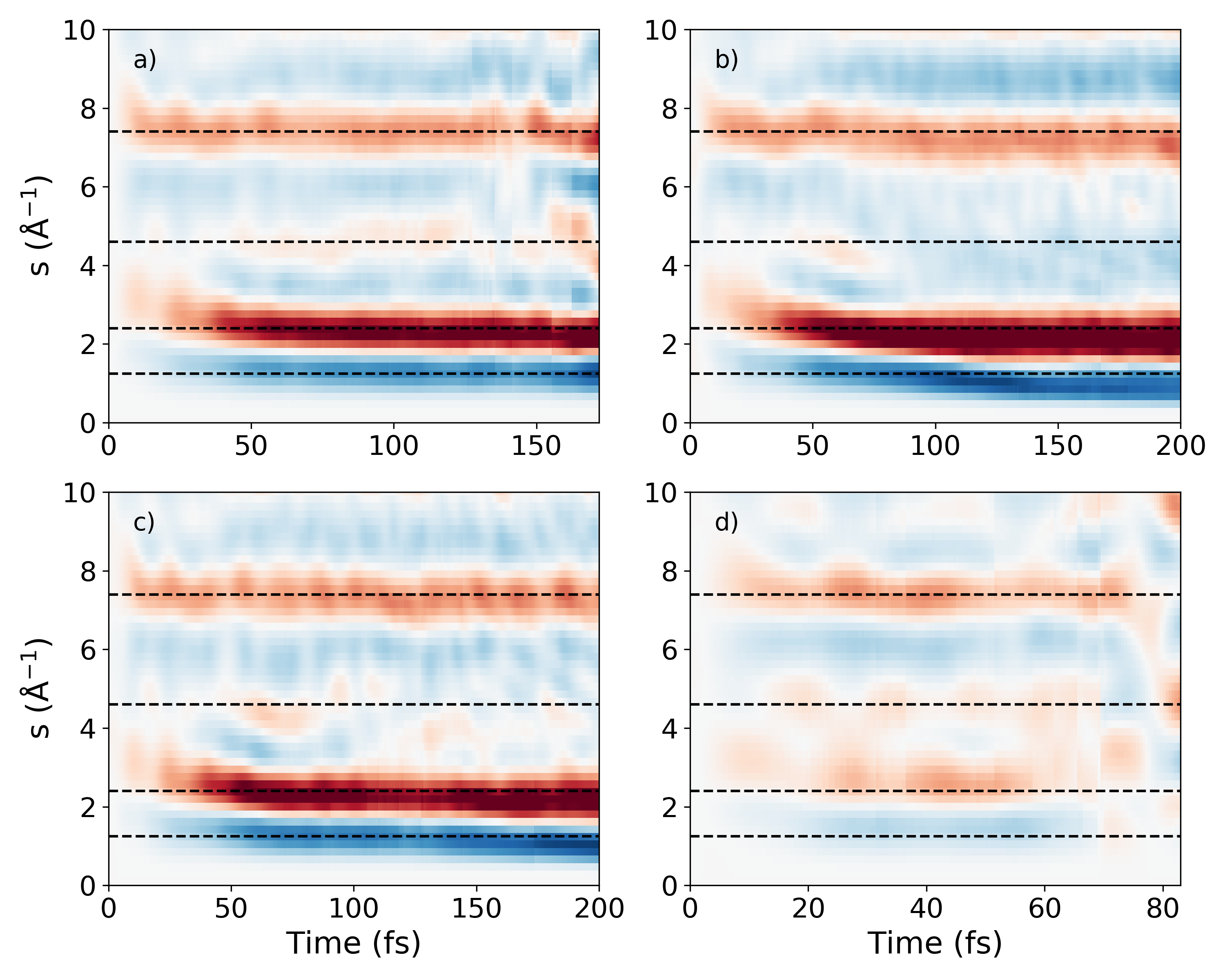}
    \caption{Percent difference UED patterns for each cluster of trajectories reported in Fig.\ \ref{fig:bonds}. Subplots show the clustered UED for: a) $\mathrm{\alpha}$-CC cleavage, b) ethene production, c) \ce{CO} production, and d) FC. Clusters that do not contain data until the target 200 fs endpoint are truncated at the final data point.} %\YYY{FIGURE NEEDS LARGER FONT FOR TEXT. COULD GET RID OF x-axis numbers on top row and y-axis numbers in second column.}}
    \label{fig:ued_clusters}
\end{figure}

 UED cross-section percent differences shown in Fig.\ \ref{fig:ued_full} present four distinct features at $s\approx$ 1.25, 2.4, 4.6, and 7.4 \AA$^{-1}$. These features constitute a fingerprint of the structural dynamics taking place in cyclobutanone, therefore, they can be interpreted with the help of additional static signals for the key geometries discussed in Sections \ref{sec: pathways} (purple line in Fig.\ \ref{fig:static_signal}). A static signal for a fixed time (80 fs for panels a and b, and 150 fs for panels c and f) is also shown in Fig.\ \ref{fig:static_signal} for each pathway (green lines), showing some broadening of the peaks due to vibrational dispersion, along with subtle shifts of peak maxima. In addition, the electron scattering cross-sections are calculated for the four individual outcomes using groups of trajectories in the dynamics simulation (\emph{i,ii,iii,iv} in Section \ref{sec: pathways}) and shown in Fig.\ \ref{fig:ued_clusters}, to aid the interpretation of the results. Due to the significant similarities between the photoproducts and the high degree of symmetry in cyclobutanone, there is a large number of overlapping features in all static and individual cluster signals (see Figures \ref{fig:static_signal} and \ref{fig:ued_clusters}). Based on these similarities, we postulate that the strong negative signal at $\sim$1.25 \AA$^{-1}$ and the strong positive feature at $\sim$2.4 \AA$^{-1}$, common to all possible photoproducts,  are related to the breaking of one or two $\mathrm{\alpha}$-bonds to produce open-ring (\emph{i}), \ce{CO} and \ce{CH_2CH_2CH_2} (\emph{ii}), or ethene and ketene (\emph{iii}). These two features, which reach a maximum at $t\sim50$ fs, are then related to the electronic state population transferring through the S$_2$/S$_1$ CI and subsequently through the S$_1$/S$_0$ CI (see Figures \ref{fig:static_signal}d), \ref{fig:static_signal}e) and \ref{fig:static_signal}f)). This timescale is consistent with what is observed in the populations of S$_1$ and the S$_0$ ground state as shown in Fig.\ \ref{fig:Population}, but significantly faster than that observed in reference \citenum{Kuhlman2012a}. It is possible that our simulations underestimate the time in S$_2$ due to a smaller barrier, as discussed in the context of Fig.\ \ref{fig:LIIC} for SA(3)-CASSCF(12,12) compared to the matching XMS-CASPT2, resulting in somewhat fast decay and product formation.

\par The other UED feature common to the total signal (Fig.\ \ref{fig:ued_full}) and all individual pathways (Fig.\ \ref{fig:ued_clusters}) is an oscillatory positive signal at $s\approx7.4$ \AA$^{-1}$. For the initial 50 fs of simulation, in this $s$ region, one can see strong coherent motions with an oscillatory period of $\sim15$ fs, indicating a possible coherent vibrational motion. All bonds between heavy atoms (C and O) are plotted for an exemplary trajectory in the SI (Fig.\ 6). Only the \ce{CO} bond length has a period that matches these oscillations, suggesting the carbonyl stretch may be responsible for this feature. 

\par  An additional feature can be observed at $s\sim4.6$ \AA$^{-1}$ with weak positive intensity for the initial $80$ fs of the time-resolved UED, becoming weakly negative after. This positive weak feature ($4-6$ \%) only shows in those pathways where $\beta$-bond breaking does not occur (Figures \ref{fig:ued_clusters}a) and \ref{fig:ued_clusters}c)) and it is negative in the one where this happens (Fig.\ \ref{fig:ued_clusters}b)). When looking at the static signals in Fig.\ \ref{fig:static_signal}, one can see that the \ce{CO} dissociation channel (Fig.\ \ref{fig:static_signal}f)) yields a positive signal at $s\sim4.6$ \AA$^{-1}$, and that this feature is also present in the ring-open structures (Fig.\ \ref{fig:static_signal}c) and \ref{fig:static_signal}d)) at $s\sim4.9$ \AA$^{-1}$. Therefore, one can relate the sign and strength of this feature to the separation between the two carbons forming the $\beta$ bonds in the molecule (see inset in Fig.\ \ref{fig:bonds}). The change of sign for this feature at $t\approx80$ fs, together with the great influence path \emph{iv)} has on the dynamics, suggests the ethene and ketene formation can be directly observed by looking at this region of the time-resolved UED signal.

\par It is important to note that the FC group presents oscillatory behavior in all of the aforementioned UED features. These oscillations can be explained by taking into account that the molecule is vibrationally excited ({\it{i.e.}} hot) when the FC region is accessed, giving rise to high-amplitude motions. This behavior is not seen in the averaged signal due to the small contribution of this feature to the overall dynamics.

%\par Finally, we can observe the constant feature at 7.4 \AA$^{-1}$. For the initial 50 fs of our UED pattern, there are strong coherent motions with an oscillatory period of $\sim$15 fs indicating a strong coherent vibrational motion from a bond. All bonds between heavy atoms (non-hydrogen) are plotted for an exemplary trajectory in SI XXX. Only the \ce{CO} bond length has a period that matches these oscillations, suggesting carbonyl stretch present in all structures is responsible for this feature. Figures \ref{fig:static_signal} and \ref{fig:ued_clusters} reveal that the feature is present in all structures with similar intensities hence making the feature constant intensity throughout the UED.

%\par \YYY{It is important to note that the FC group presents oscillatory behaviour in all of the aforementioned UED features. These oscillations can be explained by taking into account that the molecule is vibrationally excited when the FC region is accessed after the excitation takes place, giving rise to higher amplitudes. This behaviour can not be observed in the averaged signal due to the small contribution of this feature to the overall dynamics. }

%\par We have, therefore, simulated the time-resolved gas-phase UED for cyclobutanone using multi-state MASH. Data required to replot the total UED (Figure \ref{fig:ued_full}) is given in XXX for easy comparison to other theory and experimental signals. 

\section{Conclusions \label{sec:conclusion}}

\par In summary, motivated by an experiment at the SLAC MeV-UED facility and the associated prediction challenge, we have applied multistate MASH and SA(3)-CASSCF(12,12)/aug-cc-pVDZ to identify the gas-phase photodynamics of cyclobutanone after photoexcitation by 200 nm pump pulses into the n-3s Rydberg state. Mechanistic details have been identified using both a static analysis and nonadiabatic dynamics, finding a fast decay from the S$_2$ state into the S$_1$ via two CIs. A further two CIs allow for similarly fast access to the ground state. This results in the same set of photoproducts previously observed, namely $\mathrm{\alpha}$ ring-opening and ethene+ketene production, the former of which can further dissociate on the ground electronic state to liberate \ce{CO}.

\par For direct comparison with experimental results we have also predicted the gas-phase isotropic UED signal of cyclobutanone (see Fig.\ \ref{fig:ued_full}). We find that there is significant overlap between several of the products meaning that the UED signal contains multiple features. However, several distinct features indicate the formation of reaction products and loss of cyclobutanone in the equilibrium geometry. The timescale of the reaction can also be inferred from the UED signals, which, unsurprisingly, closely matches that of our simulated reaction dynamics. However, we do note that the reaction, in our simulations, proceeds significantly faster than previous observations.\cite{Kuhlman2012a, Kuhlman2012}  

\par Significant caveats remain for our predictions. The instability of the active space means that we could be potentially blind to certain reaction channels. Additionally, the value obtained for each channel's quantum yield is susceptible to significant errors from the choice of electronic structure method and the associated instabilities. This does highlight the need for robust 'black-box' electronic structure methods compatible with {\it{on-the-fly}} nonadiabatic dynamics simulations. Methods such as ADC(2) or TDDFT already exist, but are not universally applicable as they have well-documented flaws. A potential way forward, could be the selected CI family of methods,\cite{Tong2000, Coe2013} some benchmarking of which has been already been included in our results.  Further development and benchmarking of these methods is required to confirm whether they indeed offer accurate electronic structure in a more robust, black-box fashion. Also, it is important to add that the truncation of trajectories due to uncoupled hops affects the statistics of the dynamics, and therefore needs to be refined in future work. Finally, the multistate MASH used in the simulations is still being developed, and we anticipate that refinement of the algorithm will improve the results further.
%our approach of dealing with uncoupled hops needs to be refined in the future.
%needs to be corrected in the future. 
\section{Acknowledgements}
The authors would like to thank the organizers of the cyclobutanone prediction challenge for arranging this special edition. AK, MJP, LH, and AP acknowledge funding from the Leverhulme Trust (RPG-2020-208), and AK, AMC and MJP further acknowledge EPSRC EP/V006819. AK also acknowledges EPSRC EP/V049240 and grant DE-SC0020276 from the US Department of Energy. JER was funded by a mobility fellowship from the Swiss National Science Foundation under Award P500PN\_206641/1.

%\nocite{*}
\bibliography{CYCLOBUTANONE}% Produces the bibliography via BibTeX.

\end{document}